\documentclass[conference]{IEEEtran}
\IEEEoverridecommandlockouts
\usepackage{cite}
\usepackage{amsmath,amssymb,amsfonts}
\usepackage{algorithmic}
\usepackage{algorithm}
\usepackage{algorithmic}
\usepackage{graphicx}
\usepackage{textcomp}
\usepackage{xcolor}
\def\BibTeX{{\rm B\kern-.05em{\sc i\kern-.025em b}\kern-.08em
    T\kern-.1667em\lower.7ex\hbox{E}\kern-.125emX}}
\begin{document}

\title{CRB-Guided Framework Design and Resource Allocation for Indoor mmWave ISCC Systems}
\author{\IEEEauthorblockN{Zhonghao Liu\IEEEauthorrefmark{1}, Yahao Ding\IEEEauthorrefmark{1}, Yinchao Yang\IEEEauthorrefmark{1}, and Mohammad Shikh-Bahaei\IEEEauthorrefmark{1}}
\IEEEauthorblockA{\IEEEauthorrefmark{1}King's College London\\
Email: \{zhonghao.liu, yahao.ding,  yinchao.yang, m.sbahaei\}@kcl.ac.uk}}

\maketitle

\begin{abstract}
Integrated sensing, communication, and computation (ISCC) provides a promising framework for indoor human-centric applications. In these applications, short-term human pose prediction facilitates continuous human tracking and resource allocation in advance. In this paper, we propose a Cramér–Rao bound (CRB) guided resource allocation framework for indoor mmWave ISCC systems to minimize the human pose prediction error under communication, latency, and energy constraints. 
We characterize the impact of sensing power on range-estimation uncertainty and point-cloud perturbation based on the CRB. To capture the impact of computation resources on prediction performance, we adopt an adaptive-depth Mamba-based pose prediction model, where lightweight prediction heads are attached after every layer to enable inference with different model depths. With this unified sensing-computation modeling, we establish a quantitative relationship among sensing power, model depth, and prediction error. Furthermore, we formulate a joint resource allocation problem to minimize the pose prediction error. To solve this problem efficiently, we develop an alternating optimization (AO)-based algorithm, where closed-form solutions are derived for the sensing power and model depth update steps. Simulation results show that the proposed scheme significantly reduces pose prediction error compared with baseline methods, validating its effectiveness for resource-constrained indoor human-centric ISCC systems.
\end{abstract}

\begin{IEEEkeywords}
ISCC, pose prediction, mmWave, resource allocation.
\end{IEEEkeywords}

\section{Introduction}
Integrated sensing, communication, and computation (ISCC) has been regarded as a promising framework for next-generation intelligent wireless networks, since it enables wireless systems to improve resource efficiency by sharing spectrum and hardware resources \cite{10812728}. With the development of intelligent wireless networks in indoor environments, sensing tasks are gradually evolving from conventional localization and detection toward human-centric applications, such as human motion monitoring and activity recognition \cite{chen2025sensing}. Short-term human pose prediction plays an important role in such human-centric applications \cite{fujita2023future}. Specifically, the predicted positions of human joints can provide a prior estimate of the spatial structure and location of the human body. With this prior information, the access point (AP) can narrow the beam scanning region, select candidate sensing beams, and allocate resources in advance. In this context, millimeter-wave (mmWave) sensing has attracted increasing attention for indoor human-centric sensing tasks, as it can provide privacy-preserving spatial observations from reflected radio signals \cite{su2025high}.

Existing studies on ISAC/ISCC resource allocation have mainly focused on physical-layer performance from the perspective of wireless resource optimization. In \cite{liu2023snr}, Liu et al. considered a RIS-assisted ISAC system with self-interference and maximized the communication sum rate under signal to noise ratio (SNR) and Cramér–Rao bound (CRB) constraints. In \cite{li2024maximizing}, Zhang et al. jointly optimized the beamforming matrix and power to maximize communication sum rate while satisfying CRB constraints. Ren et al. investigated the CRB-rate tradeoff in multi-antenna multicast ISAC channels by optimizing the transmit covariance matrix \cite{10008661}. 
For indoor human sensing, mmWave radar has also been widely explored. Specifically, mmHPE was proposed in~\cite{10707266} to improve mmWave point-cloud quality through target boundary enhancement and perform multi-scale human pose estimation. Tang et al. proposed GF-DecNet in~\cite{tang2025gf}, which combines geometry-aware feature extraction with biomechanical constraints for robust human pose estimation.

However, physical-layer metrics cannot directly reflect the final performance of human sensing tasks. Moreover, existing learning-based mmWave human sensing methods usually focus on network architecture design under fixed sensing configurations.
When the sensing power is reduced due to the competition from communication and computation, the received echo SNR decreases, leading to larger range-estimation uncertainty, which is further manifested as point-cloud jitter \cite{11364125}. This perturbation can degrade pose prediction accuracy.
Therefore, a quantitative link between physical-layer sensing quality and task-level performance is still lacking.

Beyond sensing uncertainty, the prediction performance is also affected by the computation resources. A deeper spatio-temporal
network usually improves prediction performance \cite{jain2016structural},  but increases computation energy and inference latency \cite{ding2026energy}. Under strict energy or delay constraints with a fixed computation frequency, full-layer inference may not be feasible. Therefore, the inference depth can be treated as a controllable computation resource to characterize the tradeoff between prediction accuracy and
computation cost.
Motivated by the above limitations, we consider an indoor mmWave ISCC framework that combines CRB-guided sensing uncertainty modeling with adaptive-depth pose prediction.
The main contributions of this paper are summarized as follows:
\begin{itemize}
    \item We establish a quantitative relationship among sensing power, model depth, and pose prediction error. 

    \item We develop an adaptive-depth Mamba-based pose prediction model to enable flexible inference under energy and latency constraints.

    \item We formulate a joint resource allocation problem to minimize pose prediction error, and develop an AO-based algorithm with closed-form updates for efficient optimization.
\end{itemize}

\section{System Model}
\begin{figure}[!t]
    \centering
    \includegraphics[width=0.45\textwidth]{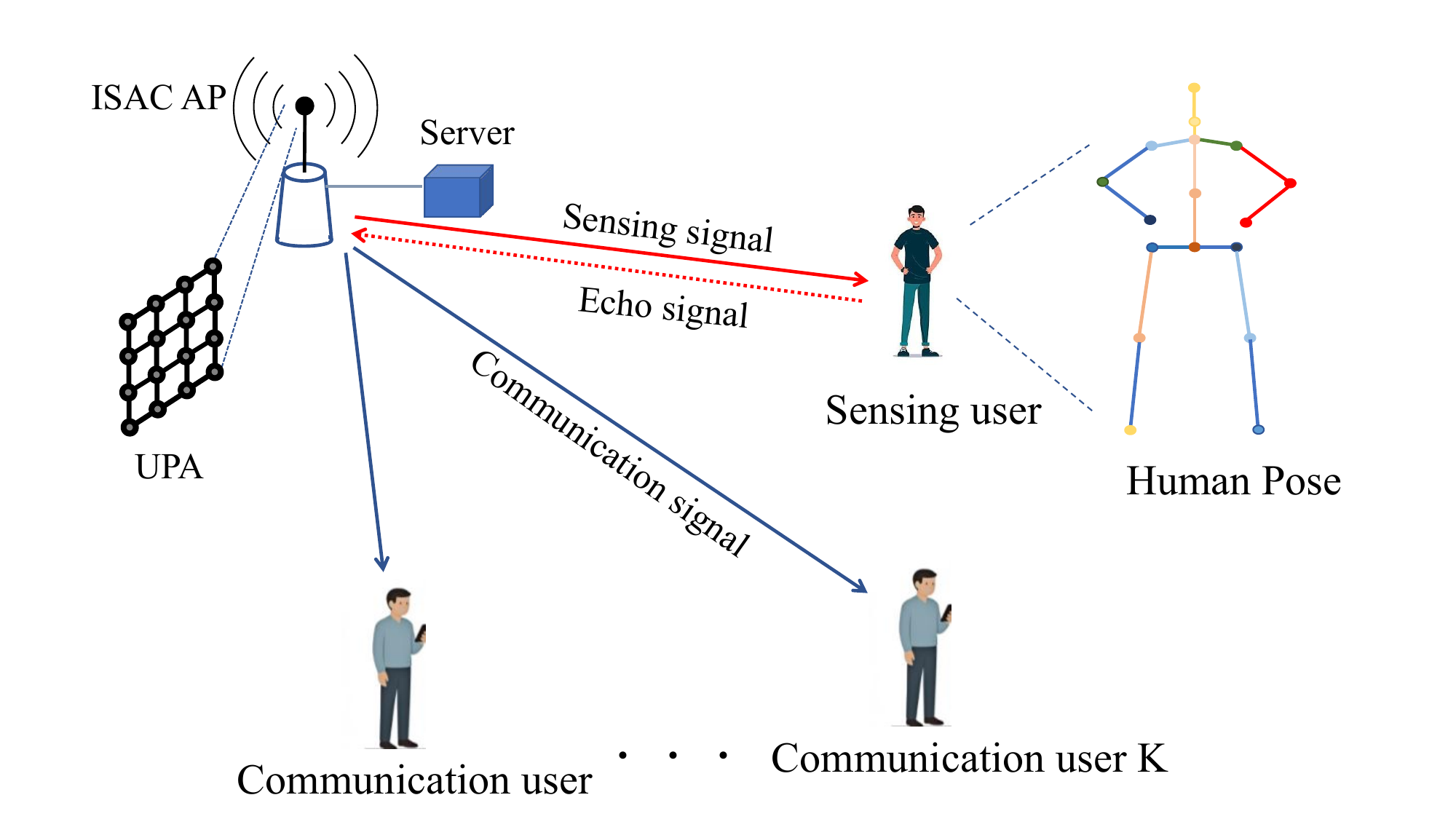}  
    \caption{Illustration of the considered indoor mmWave ISCC system for human pose prediction and tracking.}
    \label{fig:1}
\end{figure}
As shown in Fig.~\ref{fig:1}, we consider an indoor mmWave ISCC system under limited resources where an AP equipped with a uniform
planar array (UPA) consisting of $N_t=N_xN_z$ antenna elements serves $K$ communication users and performs human sensing simultaneously. Frequency division is adopted to avoid interference between communication and sensing. At time slot $t$, the transmitted signal can be represented as:
\begin{equation}
\mathbf{x}[t] =  \sum_{k=1}^{K}\,\sqrt{p_k}\mathbf{w}_{c,k}\mathbf s_{c,k}[t] +  \sqrt{p_r}\mathbf{w}_r \mathbf{s}_r[t],
\end{equation}
where $p_k$ and $p_r$ denote the communication power of user $k$ and the sensing power, respectively. $\mathbf{s}_{c,k}[t]$ denotes the communication symbol for user $k$ at time slot $t$, satisfying $\mathbb{E}\!\left[\mathbf{s}_{c,k}[t]\mathbf{s}_{c,k}^H[t]\right]=1$. Similarly, $\mathbf s_r[t]$ denotes the sensing waveform satisfying $\mathbb{E}\!\left[|\mathbf s_r[t]|^2\right]=1$. Communication and sensing signals are statistically independent, satisfying $\mathbb{E}[\mathbf s_{c,k}[t]\mathbf s_r^*[t]]=\mathbf 0$, $\forall k$.  $\mathbf{w}_{c,k}\in \mathbb{C}^{N_t}$ and $\mathbf{w}_r\in \mathbb{C}^{N_t}$ denote the communication beamforming vector for user $k$ and the sensing beamforming vector, respectively. Specifically, $\mathbf{w}_r$ is selected from a predefined unit-norm beam codebook $\mathcal{B}=\left\{\mathbf{a}(\theta_i,\phi_j)\right\}$, where $(\theta_i,\phi_j)$ denotes a candidate azimuth-elevation direction. For the considered UPA employed in the Oxz plane, $\mathbf{a}(\theta_i,\phi_j)$ is given by:
$\mathbf{a}(\theta_i,\phi_j)=\mathbf{a}_x(\theta_i,\phi_j)\otimes \mathbf{a}_z(\theta_i,\phi_j)$,
where the horizontal steering vector $\mathbf{a}_x(\theta_i,\phi_j)$ can be given by $\left[1, e^{j \frac{2\pi d}{\lambda}\sin(\phi_j)\cos(\theta_i)}, \ldots, e^{j (N_x-1)\frac{2\pi d}{\lambda}\sin(\phi_j)\cos(\theta_i)}\right]^T$, and the vertical steering vector $\mathbf{a}_z(\theta_i,\phi_j)$ is denoted by $\left[1, e^{j\frac{2\pi d}{\lambda}\sin(\phi_j)}, \ldots, e^{j(N_z-1)\frac{2\pi d}{\lambda}\sin(\phi_j)}\right]^T$. 

The communication beamforming vector is designed using zero-forcing (ZF) precoding to suppress inter-user interference. Thus, the received signal of communication user $k$ at time slot $t$ can be written as:
\begin{equation}
\label{comxinhao}
y_k[t] = \sqrt{p_k}\mathbf{h}_k^{H}\mathbf{w}_{c,k}\mathbf s_{c,k}[t] + n,
\end{equation}
where $\mathbf{h}_k \in \mathbb{C}^{N_t}$ denotes the downlink channel between the AP and user $k$, and $n\sim \mathcal{CN}(0,\sigma_n^2)$ is the additive white Gaussian noise. For notational simplicity, the time index is omitted in the following. Based on the predicted positions of human joints from the previous time slot, the AP selects $P$ candidate beams from the codebook and scans them sequentially within the current time slot. Thus, at time slot $t$, the echo signal can be given by:
\begin{align}
\mathbf{e}[t]
&= \sum_{p=1}^{P}
\beta_{p} \sqrt{p_r}
\mathbf{A}_{p} 
\mathbf{w}_{r,p} \mathbf{s}_r\!\left(t-\tfrac{2d_{p}}{c}\right) + z,
\label{eq:total_echo}
\end{align}
where $\beta_{p}$ = $\zeta / 2d_{p}$ is the path-loss coefficient, $d_{p}$ is the propagation distance, $\zeta$ is the radar cross section (RCS), $\mathbf{A}_{p}$ = $ \mathbf{a}(\theta_{p}, \phi_{p})
\mathbf{a}^H(\theta_{p}, \phi_{p})$ denotes the directional response matrix associated with the $p$-th scanned beam, $\mathbf{w}_{r,p}^{}$ is the $p$-th scanned sensing beam at time slot $t$, $c$ is the speed of light, and $z\sim\mathcal{CN}(0,\sigma_z^2)$ is the additive white Gaussian noise.
\begin{figure*}[!t]
    \centering
    \includegraphics[width=0.88\linewidth]{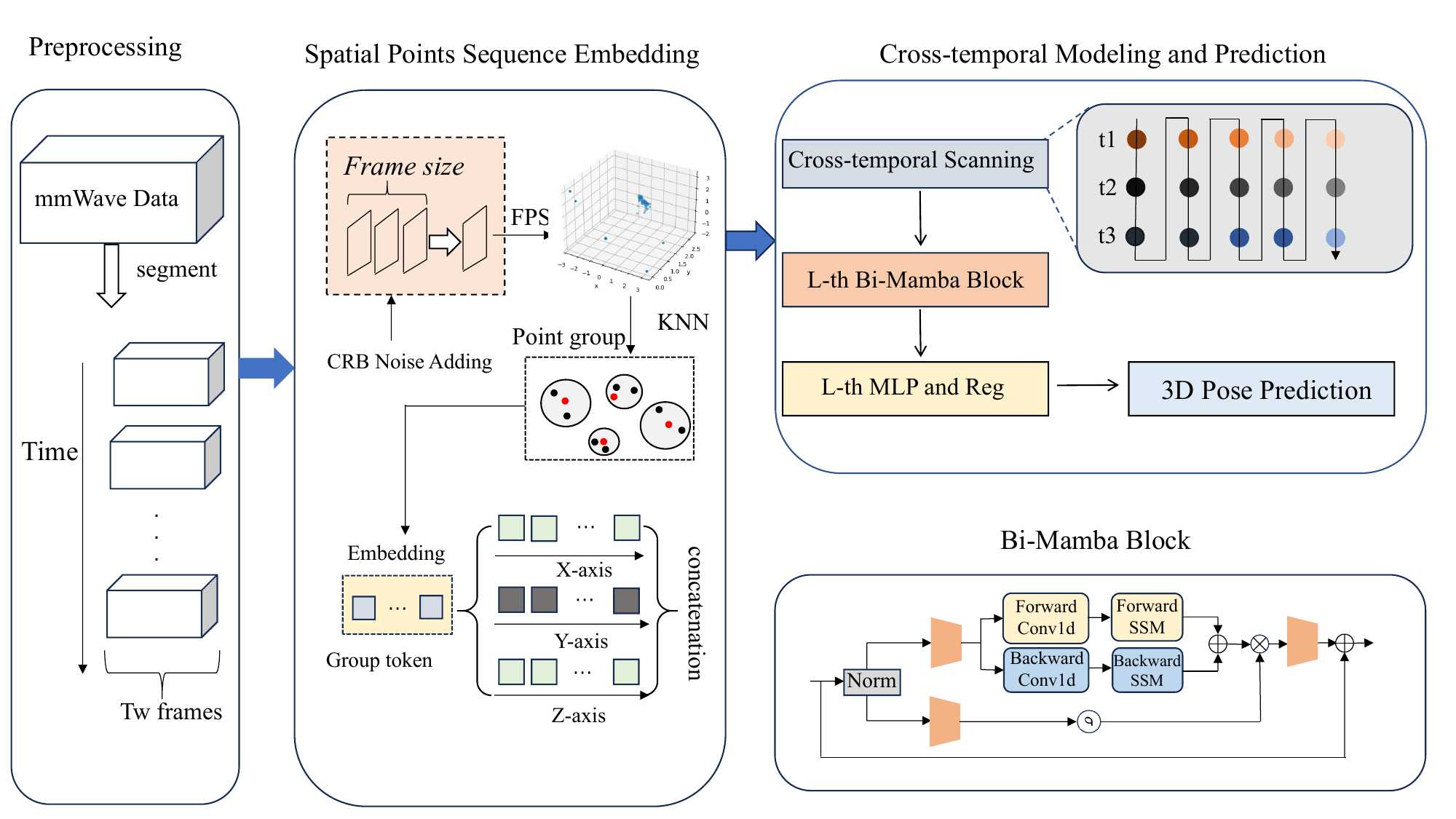}  
    \caption{Architecture of the proposed CRB-Mamba framework for mmWave point-cloud-based human pose prediction.}
    \label{fig:2}
\end{figure*}

Following \eqref{comxinhao}, the communication SNR of user $k$ is:
\begin{equation}
\mathrm{SNR}_k=
\frac{p_k\left|\mathbf{h}_k^H{\mathbf{w}}_{c,k}\right|^2}
{\sigma_n^2}.
\label{eq:SINR_fixed}
\end{equation}

Following \eqref{eq:total_echo}, the echo SNR corresponding to the $p$-th scanned beam can be denoted as:
\begin{equation}
\mathrm{SNR}_{p}
=
\frac{
\zeta^{2} p_r
}{
4d_{p}^{4}\,\sigma_{z}^{2}
}.
\label{eq:SNR_s_o}
\end{equation}
Following \cite{11364125}, the uncertainty in the estimated distance can be characterized by the CRB, which can be denoted by:
\begin{equation}
\sigma_{d,p}^{2}=\frac{c^{2}}{8\pi^{2}\mathrm{SNR}_{p}B_r^{2}},
\label{eq:CRB}
\end{equation}
where $B_r$ is the bandwidth of sensing. Based on that, the estimated distance $\tilde{d_p}$ can be denoted by $\tilde{d_p}$ = $d_p + \Delta d_p$, where $\Delta d_p\sim\mathcal N(0,\sigma_{d,p}^{2})$. The estimated distance set $\tilde{\mathbf D}=\{\tilde d_{p}\}$ can be further mapped into point clouds, where the point corresponding to the $p$-th beam is represented by $\mathbf p_p=\tilde d_{p}\,[\cos\theta_p\cos\phi_p,\ \cos\theta_p\sin\phi_p,\ \sin\theta_p]^T$. 
\section{ADAPTIVE-DEPTH MAMBA-BASED POSE PREDICTION}
When sensing power is reduced due to resource competition with communication and computation, the resulting range-estimation uncertainty is manifested as point-cloud jitter. To improve pose prediction performance under this perturbation and resource constraints, we propose an adaptive-depth CRB-Mamba framework, as shown in Fig. \ref{fig:2}. 

For point-cloud perturbation modeling, $\mathrm{SNR}_o$ is defined as the sensing SNR
of the frame center point. $\mathrm{SNR}_p$ is obtained by scaling $\mathrm{SNR}_o$ according
to its relative range, and the corresponding CRB-derived variance is calculated by \eqref{eq:CRB}.
To model the temporal dependencies efficiently, we adopt a bidirectional Mamba (Bi-Mamba) network as the backbone because its selective state-space model enables efficient long-sequence modeling with linear complexity. The discrete state-space model can be
written as
\begin{equation}
\mathbf{h}_b
=
\bar{\mathbf{A}}\mathbf{h}_{b-1}
+
\bar{\mathbf{B}}\mathbf{x}_b,
\quad
\mathbf{y}_b
=
\bar{\mathbf{C}}\mathbf{h}_b
+
\mathbf{D}\mathbf{x}_b,
\end{equation}
where $\mathbf{x}_b\in\mathbb{R}^{d}$, $\mathbf{h}_b\in\mathbb{R}^{N}$,
and $\mathbf{y}_b\in\mathbb{R}^{d}$ denote the input, hidden state,
and output at time step $b$, respectively. The matrices
$\bar{\mathbf{A}}$, $\bar{\mathbf{B}}$, $\bar{\mathbf{C}}$, and
$\mathbf{D}$ are learnable parameters.

For each frame, we first select a set of representative group centers using farthest point sampling (FPS), and then construct local neighborhoods around each center via $K$-nearest neighbors (KNN). Each neighborhood is normalized by subtracting its center to obtain relative coordinates. After serializing the original point cloud into a 1D sequence, we employ a 6-layer Bi-Mamba network for training. Considering that edge-side power consumption and delay constraints may limit the execution of full network inference, we attach a lightweight prediction head after each Bi-Mamba layer, enabling the model to still output the corresponding pose prediction results when full-layer inference is infeasible.
During training, the original action sequence is partitioned into overlapping samples by a sliding window of length $T_w$. Each clip is used as one training sample, and the model predicts
the 3D joint coordinates of the next frame. The network is optimized with Adam. Each frame contains $J$ joints, and the predicted and
ground-truth coordinates of joint $j$ in the $cp$-th clip are denoted by
$\hat{\mathbf{p}}^{(cp)}_{j}$ and $\mathbf{p}^{*(cp)}_{j}$, respectively.
\begin{equation}
\mathcal{L_{MSE}}=\frac{1}{N_bJ}\sum_{}^{N_b}\sum_{}^{J}\bigl\|\hat{\mathbf{p}}^{(cp)}_{j}-\mathbf{p}^{*(cp)}_{j}\bigr\|_2^2,
\end{equation}
where $N_b$ is the batch size.

\section{PROBLEM FORMULATION AND SOLUTION}

\subsection{Problem Formulation}
The energy consumption of the AP consists of three parts: sensing transmission energy $E_r$, communication transmission energy $E_c$, and local computation energy $E_{comp}$. 
Let $C_{pc}$ denote the CPU cycles required for point-cloud generation in one time slot, $C_{b}$ represent the layer-independent CPU cycles required for feature construction and the final prediction head, and $C_L$ denote the CPU cycles required by each inference layer. Thus, the computation latency and energy can be expressed as:
\begin{equation}
\tau_{\mathrm{comp}}
=\frac{C_{pc}+C_{\mathrm{b}}+L\,C_L}{f},
\label{eq:tau_comp}
\end{equation}
\begin{equation}
E_{comp} = \gamma \big(C_{pc}+C_{\mathrm{b}}+L\,C_L\big)f^2,
\end{equation}
where $f$ is the fixed computation frequency. Let $m(p_r,L)$ denote the prediction error, the optimization problem is given by:
\begin{subequations}\label{eq:simplified_problem}
\begin{align}
\min_{p_r,\,L,\{p_k\}_{k=1}^{K}}\quad & m(p_r,L) 
\label{eq:simplified_problem_obj}\\
\text{s.t.}\quad 
& T_0 + \tau_{\mathrm{comp}} \le T, 
\label{eq:simplified_problem_delay}\\
& E_c + p_r T_0 + E_{\mathrm{comp}} \le P_{\max}T, 
\label{eq:simplified_problem_energy}\\
& SNR_k \ge SNR_{\min}, \quad \forall k, 
\label{eq:simplified_problem_rate}\\
& \mathrm{SNR}_o \ge \gamma_{\min}^{\mathrm{det}}, 
\label{eq:simplified_problem_snr}\\
& 1 \le L \le L_{\max}, \quad L \in \mathbb{Z}_{+}, 
\label{eq:simplified_problem_depth}\\
& 0 \le p_r+\sum_{k=1}^{K}p_k \le P_{t},
\label{eq:simplified_problem_power}
\end{align}
\end{subequations}
where $T_0$ denotes the sensing duration within one time slot, and $P_{\max}$ denotes the maximum allowable average power. $\mathrm{SNR}_{\min}$ is the minimum communication SINR threshold, while $\gamma_{\min}^{\mathrm{det}}$ denotes the minimum sensing SNR required for target detectability. In addition, $P_t$ is the maximum total transmit power at the AP. Constraint \eqref{eq:simplified_problem_delay} requires that the sensing duration and the computing latency should not exceed the total slot duration. Constraint \eqref{eq:simplified_problem_energy} represents the total energy budget. Constraint \eqref{eq:simplified_problem_rate} guarantees the communication QoS and \eqref{eq:simplified_problem_snr} ensures that the sensing SNR at the target center remains above a prescribed threshold. Constraints \eqref{eq:simplified_problem_depth} and \eqref{eq:simplified_problem_power} specify the feasible domains of the optimization variables.
\subsection{Problem Solution}
Before optimizing the sensing power and the model depth, the communication power required to satisfy the QoS constraints of all users is first determined.
\subsubsection{Empirical Modeling of Prediction Error}
We use the mean per-joint position error (MPJPE) to evaluate the pose prediction performance. MPJPE measures the average Euclidean distance between the predicted 3D joint coordinates and the corresponding ground-truth joint coordinates over all joints and testing samples. A smaller MPJPE indicates more accurate pose prediction.
\begin{figure}[!t]
    \centering
    \includegraphics[width=0.45\textwidth]{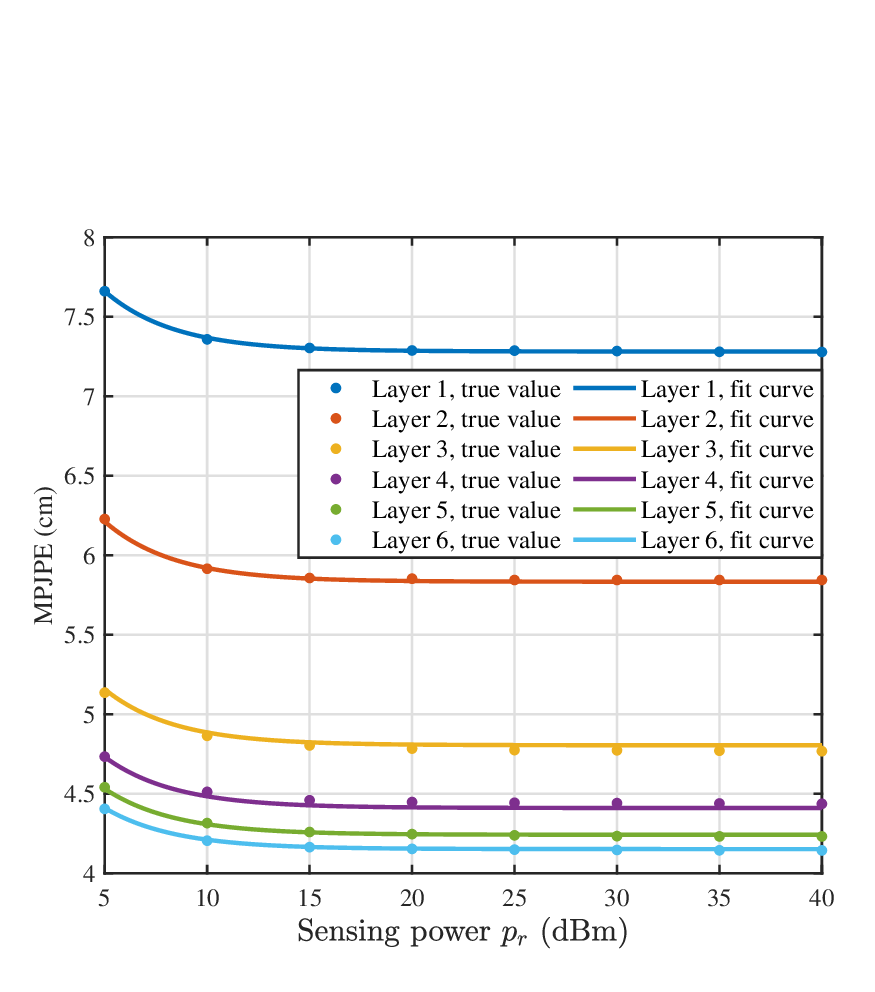}  
    \caption{Fitted MPJPE curves under different sensing power and model depth.}
    \label{fig:curve}
\end{figure}
After multiple runs and evaluations, we obtain the fitted curve shown in Fig.~\ref{fig:curve}.
The empirical results indicate that MPJPE decreases monotonically with both $p_r$ and $L$. This is because a larger sensing power reduces jitter, while a larger inference depth improves the spatio-temporal representation capability of the pose prediction model.
\subsubsection{Minimum Communication Power with Fixed Beam Directions}
The minimum communication power required to support all users is obtained by solving
\begin{subequations}\label{eq:Pc_min_problem}
\begin{align}
\min_{\{p_k\}} \quad & \sum_{k=1}^{K} p_k
\label{eq:Pc_min_problem_a}\\
\text{s.t.}\quad
&  0 \le p_k \leq P_t,\quad \forall k,\\
& \eqref{eq:simplified_problem_rate} \nonumber
\end{align}
\end{subequations}
\eqref{eq:Pc_min_problem} can be efficiently solved using CVX. Let $P_c^{\min}$ denote its optimal value. 
Hence, once the minimum communication power is obtained, the problem \eqref{eq:simplified_problem} is reduced to the joint optimization of the model depth $L$ and the sensing transmit power $p_r$, which can be expressed as:
\begin{subequations}
\begin{align}
\label{AO based method}
\min_{p_r,\,L}\quad & m(p_r,L) \\
& p_r T_0 + E_{\mathrm{comp}} \le \tilde E_{max},\label{optE}\\
&0 \le p_r \le \tilde P_{t},\label{optP}\\
&\eqref{eq:simplified_problem_delay},\eqref{eq:simplified_problem_snr},\eqref{eq:simplified_problem_depth}\nonumber,
\end{align}
\end{subequations}
where $\tilde P_{t}$ is $P_{t} - P_c^{min}$ and $\tilde E_{max}$ is $P_{max}T - P_c^{min}T$.
\subsubsection{AO-Based Algorithm}
To solve problem \eqref{AO based method}, we adopt an alternating optimization (AO)-based algorithm. Since the fitted error function $m(p_r,L)$ decreases monotonically with both $p_r$ and $L$, the original problem can be decomposed into two subproblems and solved alternately.

We first apply a continuous relaxation by relaxing the integer constraint $L \in \mathbb{Z}_{+}$ to the interval $L \in [1, L_{\max}]$. For a given sensing transmit power $p_r$, the subproblem 1 with respect to the model depth $L$ is formulated as
\begin{equation}\label{eq:subproblem_C}
\begin{aligned}
\min_{L}\quad & m(L) \\
\text{s.t.}\quad
& \eqref{eq:simplified_problem_delay}, \eqref{eq:simplified_problem_depth}, \eqref{optE}
\end{aligned}
\end{equation}
According to the latency constraint \eqref{eq:simplified_problem_delay}, we have the following upper bound on the model depth:
\begin{equation}
L_{\tau} = \frac{f(T-T_0)-C_{pc}-C_{b}}{C_L}.
\label{eq:C_bound_delay}
\end{equation}
According to the energy constraint \eqref{optE}, we have another upper bound on the model depth:
\begin{equation}
L_E= \frac{ \tilde E_{max}-p_rT_0-\gamma (C_{pc}+C_b) f^2}{\gamma C_L f^2}.
\label{eq:C_bound_energy}
\end{equation}
By combining \eqref{eq:C_bound_delay}, \eqref{eq:C_bound_energy}, and the feasible range of $L$, the optimal solution to the relaxed problem can be expressed in closed form as
\begin{equation}
L^\star=\max\left\{1,\left\lfloor \min\left\{L_{\max},L_\tau,L_E\right\}\right\rfloor\right\},
\label{eq:C_star}
\end{equation}
where $\lfloor \cdot \rfloor$ denotes the floor operation, which returns the largest integer no greater than its argument.

Similarly, for a given model depth $L$, the subproblem 2 with respect to the transmit power $p_r$ is formulated as
\begin{equation}\label{eq:subproblem_Pr}
\begin{aligned}
\min_{p_r}\quad & m(p_r) \\
\text{s.t.}\quad
& \eqref{eq:simplified_problem_snr},\eqref{optE}, \eqref{optP}. 
\end{aligned}
\end{equation}

According to the constraint \eqref{optE}, we have an upper bound on the sensing transmit power:
\begin{equation}
P_E = \frac{\tilde E_{max}-E_{\mathrm{comp}}}{T_0}.
\label{powerE}
\end{equation}

In addition, according to the sensing detectability constraint \eqref{eq:simplified_problem_snr}, the sensing power should also satisfy the following lower bound:
\begin{equation}
P_r^{\min} =
\frac{4d_p^4\sigma_z^2\,\gamma_{\min}^{\mathrm{det}}}{\zeta^2}.
\label{eq:Pr_bound_snr}
\end{equation}

By combining \eqref{powerE}, \eqref{eq:Pr_bound_snr}, and the feasible range of $p_r$, the optimal sensing power can be expressed in closed form as
\begin{equation}
p_r^{\star}=\min\left\{\tilde{P}_{t},\,P_E\right\},
\quad
\text{subject to } p_r^{\star}\ge P_r^{\min}.
\label{eq:Pr_star}
\end{equation}

With the closed-form updates of $L$ and $p_r$, the proposed method can be implemented through an AO-based iterative procedure, as summarized in Algorithm~\ref{alg:ao}. The minimum communication power is computed only once, while both the depth update and the sensing power update have closed-form solutions. Therefore, if the AO procedure converges in $I$ iterations, the overall complexity is $O(K^3+I)$.

\begin{algorithm}[t]
\caption{AO-Based Joint Optimization Algorithm}
\label{alg:ao}
\begin{algorithmic}[1]
\REQUIRE System parameters, $I_{\max}$, and convergence tolerance $\epsilon$.
\ENSURE Optimized communication power $\{p_k^\star\}_{k=1}^{K}$, sensing power $p_r^\star$, and model depth $L^\star$.
\STATE \textbf{Initialize} $i=0$, $L^{(0)}=1$, and $p_r^{(0)}=P_r^{\min}$, where $P_r^{\min}$ is obtained by \eqref{eq:Pr_bound_snr}.
\STATE Obtain the minimum communication power $P_c^{\min}$ by solving problem (12).
\STATE Update $\tilde{P}_{t}$ and $\tilde{E}_{\max}$.
\REPEAT
    \STATE Obtain $L^{i+1}$ by \eqref{eq:C_star}.
    \STATE Obtain $p_r^{(i+1)}$ by \eqref{eq:Pr_star}.
    \STATE Set $i=i+1$.
\UNTIL{the convergence criterion is satisfied}
\STATE \textbf{return} $\{p_k^\star\}_{k=1}^{K}$, $p_r^\star=p_r^{(i)}$, and $L^\star=L^{(i)}$.
\end{algorithmic}
\end{algorithm}

\section{Simulation And Results Analysis}
We conduct numerical simulations to evaluate the proposed model. The simulation
parameters are summarized in Table~\ref{tab:simulation_parameters}. The hidden feature dimension of the Bi-Mamba network is set to $384$,
and the window length is set to $T_w=8$. To evaluate the
effectiveness of joint sensing-computation optimization, two baseline
schemes are considered. The fixed $L=1$ baseline minimizes computation cost and
allocates more resources to improve sensing quality, while the
fixed $p_r=P_r^{\min}$ baseline only satisfies the minimum sensing
detectability requirement and uses the remaining resources to select
the largest feasible inference depth.
\begin{table}[t]
\centering
\caption{Simulation Parameters}
\label{tab:simulation_parameters}
\renewcommand{\arraystretch}{1.1}
\begin{tabular}{l l}
\hline
\textbf{Parameter} & \textbf{Value} \\
\hline
Number of terminals, $K$ & 4 \\
UPA size, $(N_x,N_z)$ & $(4,4)$ \\
Noise power, $\sigma_n^2$, $\sigma_z^2$ & $10^{-6}$ \\
Sensing bandwidth $B_r$ & $0.5\,\mathrm{GHz}$ \\
Minimum communication SNR, $SNR_{\min}$ & $5\,$ \\
Slot duration, $T$ & $0.1\,\mathrm{s}$ \\
Transmission duration, $T_0$ & $0.05\,\mathrm{s}$ \\
Minimum sensing SNR threshold, $\gamma_{\min}^{\mathrm{det}}$ & -5dB \\
Computation frequency, $f$ & $100\,\mathrm{MHz}$ \\
Maximum Bi-Mamba depth, $L_{\max}$ & 6 \\
Effective switched-capacitance coefficient, $\gamma$ & $10^{-25}$ \\
CPU cycles for point-cloud generation, $C_{pc}$ & 128*256*8 \\
Layer-independent CPU cycles, $C_{b}$ & 384*96*8\\
CPU cycles per Bi-Mamba layer, $C_{L}$ & 384*96*8*4 \\
\hline
\end{tabular}
\end{table}
\addtolength{\topmargin}{0.05in}
\begin{figure}[!t]
    \centering
    \includegraphics[width=0.5\textwidth]{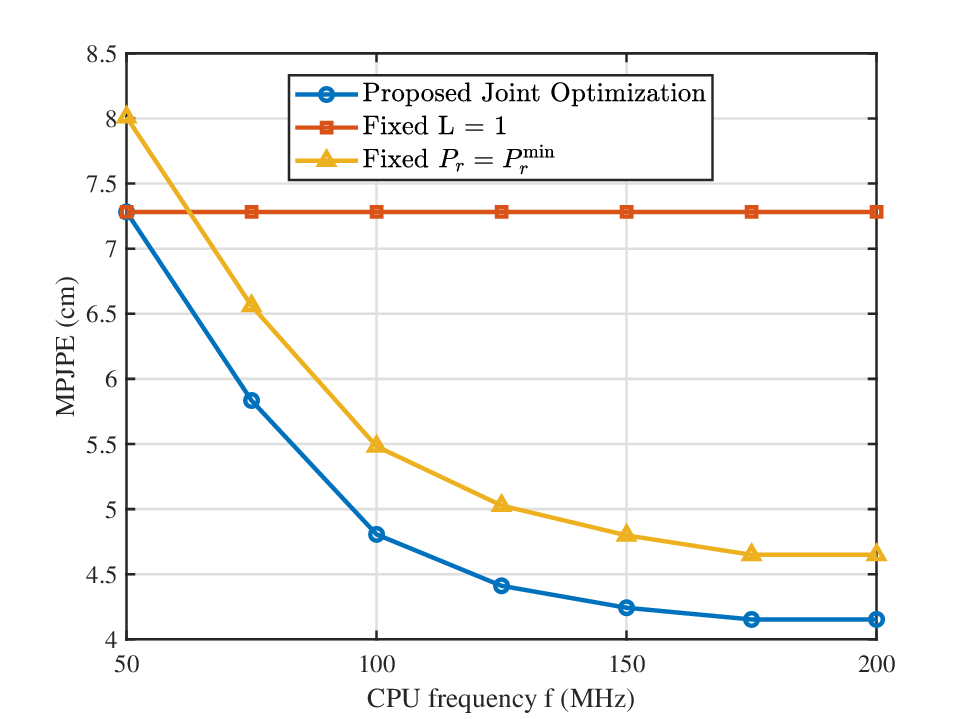}  
    \caption{MPJPE versus CPU frequency.}
    \label{figf}
\end{figure}
Fig.~\ref{figf} shows the MPJPE performance versus the CPU frequency
$f$. As $f$ increases, the proposed method significantly reduces
the MPJPE, especially in the low-frequency region, because a higher
CPU frequency relaxes the latency constraint and enables a deeper
inference model. When $f$ exceeds around 150 MHz, the performance
gradually saturates, since the model depth approaches its upper
limit. Compared with the fixed $L=1$ baseline, the proposed method
reduces the MPJPE by about $43\%$ in the high-frequency region.
Compared with the fixed $p_r=P_r^{\min}$ baseline, it achieves about
$10\%$--$13\%$ improvement, showing the benefit of jointly optimizing
sensing power and model depth.

\begin{figure}[!t]
    \centering
    \includegraphics[width=0.5\textwidth]{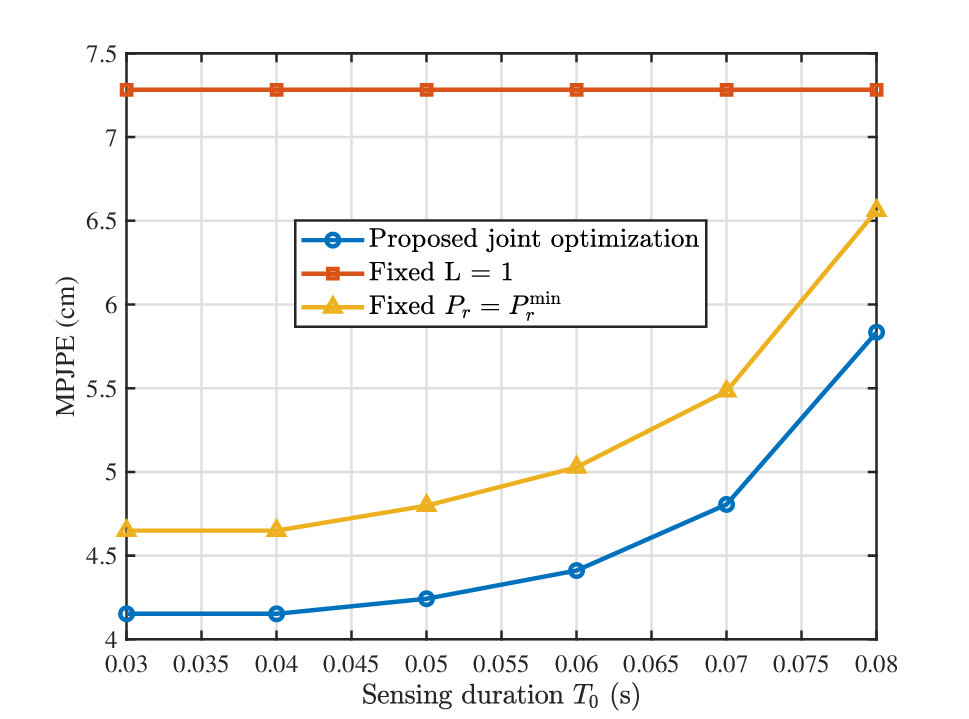}  
    \caption{MPJPE versus sensing duration $T_0$. }
    \label{figT}
\end{figure}

Fig.~\ref{figT} presents the MPJPE performance under different sensing
durations $T_0$. The proposed method maintains a low MPJPE when
$T_0$ is moderate, while the error increases when $T_0$ becomes large.
This is because a larger $T_0$ reduces the available computation
time $T-T_0$ and tightens the energy budget. Compared with the fixed
$L=1$ baseline, the proposed method achieves about $43\%$ MPJPE
reduction when $T_0$ is moderate and still maintains about $20\%$
improvement when $T_0=0.08$~s. Compared with the fixed
$p_r=P_r^{\min}$ baseline, it consistently achieves about
$10\%$--$12\%$ improvement.

\begin{figure}[!t]
    \centering
    \includegraphics[width=0.5\textwidth]{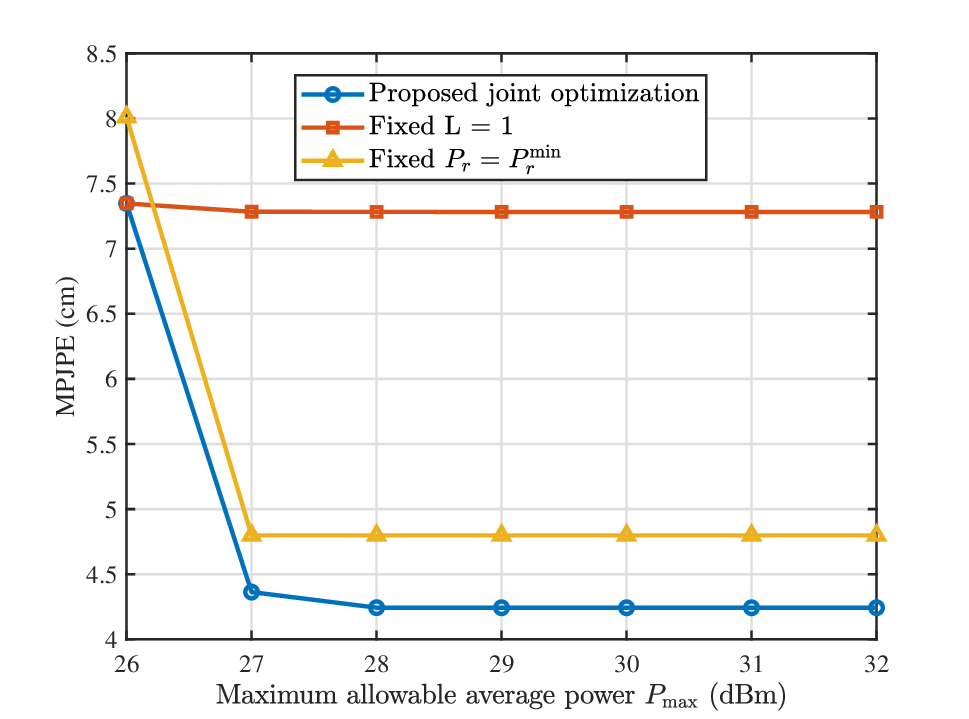}  
    \caption{MPJPE versus maximum allowable average power $P_{max}$.}
    \label{figE}
\end{figure}

Fig.~\ref{figE} illustrates the impact of the maximum allowable average
power on MPJPE. When the power budget is limited, all schemes
suffer from high prediction error due to insufficient sensing and
computation resources. As the power budget increases from 26~dBm
to 27~dBm, the proposed method sharply reduces the MPJPE from
above 7~cm to about 4.4~cm, and then gradually saturates after around
28~dBm. Compared with the fixed $L=1$ baseline, the proposed
method reduces the MPJPE by about 42\% in the medium-to-high
power region. Compared with the fixed $p_r=P_r^{\min}$ baseline,
it achieves about 10\%--12\% improvement, confirming the effectiveness
of joint sensing-computation resource allocation.

\section{Conclusion}
In this paper, we proposed a human-centric sensing framework for indoor mmWave ISCC systems. The proposed framework jointly considers CRB-guided sensing uncertainty and model-depth selection to improve human pose prediction performance under limited resources. We established a quantitative relationship among sensing power, model depth, and pose prediction error. Based on the relationship, we formulated a joint resource allocation problem to minimize the pose prediction error. To solve this problem, we developed an AO-based joint optimization algorithm, where closed-form update rules were derived for efficient optimization. Simulation results demonstrate that the proposed method effectively reduces pose prediction error compared with baseline schemes. Future work will extend the framework to multiple sensing users in diverse indoor scenarios with more complex human motion patterns.

\bibliographystyle{ieeetr}
\bibliography{main}

\end{document}